\documentclass[epsfig,prl,twocolumn,amssymb]{revtex4}

\usepackage{graphicx}

\begin{document}
\title
{
Soliton creation during a Bose-Einstein condensation
}
\begin{abstract}
We use stochastic Gross-Pitaevskii equation to study dynamics of Bose-Einstein condensation. 
We show that cooling into a Bose-Einstein condensate (BEC) can create solitons with density given by the cooling rate and by 
the critical exponents of the transition. Thus, counting solitons left in its wake should allow one to determine the critical exponents $z$ and $\nu$  for a BEC phase transition. The same information can be extracted from two-point correlation functions.
\end{abstract}
\author{Bogdan Damski and Wojciech H. Zurek}
\affiliation
{
Theoretical Division, Los Alamos National Laboratory, MS-B213, Los Alamos, NM 87545, USA
}
\maketitle

When temperature of a bosonic cloud is lowered, atoms undergo a thermal gas -- Bose-Einstein condensate phase transition. Sufficiently slow cooling process creates a condensate in the ground state. Its macroscopic quantum properties have been studied over the last decade. Faster 
cooling should result in creation of an excited condensate that contains vortices and solitons. 

Non-equilibrium condensation is an example of dynamics of phase transitions.
Their study started with investigations of 
topological defect creation in 
cosmological phase transitions \cite{kibble}. It was then pointed out that  condensed matter systems (superconductors, liquid 
$^3$He and $^4$He, etc.) can be used to test this cosmological scenario, and that the nature of the transition will determine density of defects \cite{zurek}. Kibble-Zurek 
mechanism (KZM) describes how topological defects (vortices, monopoles, kinks, etc.)
are  created during non-equilibrium second order 
phase transitions (see \cite{eksperymenty,anderson} for experimental evidence).

Data on spontaneous creation
of solitons and vortices during Bose-Einstein condensation are still 
scarce.
Two groups have seen them: 
the Anderson group observed  vortices  in a three dimensional 
trap \cite{anderson}, while
the Engels' group reported observation of solitons 
in a quasi  one dimensional (1D) setup that does not support topological
defects \cite{engels}.
Therefore, it is important to understand non-equilibrium 
dynamics of  Bose-Einstein condensation.
Moreover, it is also interesting to find out if the 
KZM, studied theoretically  and experimentally in various 
physical systems
\cite{zurek,eksperymenty,anderson,nature_kurn,laguna,kibble_today,quantum},
applies to non-topological excitations (e.g., solitons).
We propose to  kill  two birds with one stone by studying 
dynamics of Bose-Einstein condensation in a quasi-one 1D system.

As the atom cloud is cooled from above the critical point,
the system adjusts adiabatically
to driving because its relaxation time is initially short.
Near the critical point, however, the relaxation time diverges 
due to critical slowing down.  
The system goes out of equilibrium
before reaching the critical point and approximately freezes out
entering the impulse stage of its dynamics.
As the condensate 
forms, its different parts choose to break the symmetry in an uncorrelated way, 
which results in creation of defects.
Their spacing  depends on the quench rate given by  
the rate of temperature change \cite{zurek}: the slower we go, the more adiabatic the
evolution is, and so the larger the separation between the defects will be.
The transition from adiabatic to impulse regime happens
when the relaxation time $\tau$ becomes comparable to the 
quench rate $\varepsilon/\dot{\varepsilon}$ ($\varepsilon$ is the 
distance from the critical point):
\begin{equation}
\tau(\varepsilon(t)) = \left|\frac{\varepsilon}{\dot{\varepsilon}}\right|.
\label{zurek_eq}
\end{equation}
Assuming a linear quench, i.e., 
$|\dot{\varepsilon}|=\tau_Q^{-1}$, 
where $\tau_Q$ is the quench timescale, the system goes out 
of equilibrium at 
\begin{equation}
\hat\varepsilon \sim \tau_Q^\frac{-1}{1+z\nu}.
\label{epsilon_hat}
\end{equation}
This imprints a characteristic time 
scale $\hat t=\tau_Q\hat\varepsilon=\tau_Q^{z\nu/(1+z\nu)}$ onto the system,
so that defect density depends on time through $t/\hat t$ or 
equivalently $\varepsilon(t)/\hat \varepsilon$. 
The typical distance between defects scales as 
\begin{equation}
\hat\xi = \xi(\hat\varepsilon) \sim \tau_Q^\frac{\nu}{1+z\nu},
\label{xi_hat}
\end{equation}
in the KZM  freeze out picture \cite{zurek}.
Above $z$ and $\nu$ are the critical exponents defining the equilibrium 
coherence length $\xi$ and  relaxation timescale $\tau$:
\begin{equation}
\xi \sim |\varepsilon|^{-\nu}, \ \tau \sim |\varepsilon|^{-z\nu}.
\label{xi_tau_eq}
\end{equation}
Thus, KZM (presented above) proposes a way to extract 
essential features of the non-equilibrium phase transition 
dynamics from equilibrium critical behavior of a system.

The exact simulation of  Bose-Einstein condensation 
is extremely involved (if at all possible) \cite{ashton}. 
We restrict ourselves to a tractable model: 
the stochastic Gross-Pitaevskii equation (SGPE)
that has been successfully  applied to studies of Bose-Einstein
condensates lately \cite{ashton}. 
Our calculations provide the first numerical results on dynamics of soliton 
production in the course of second order phase transitions,
and extend the KZM to non-topological excitations.
Former studies of  1D models, driven across a critical point, were focused on 
systems supporting topological defects (e.g.,  
kinks \cite{laguna}).

Stochastic Gross-Pitaevskii equation reads \cite{sgpe}:
\begin{equation}
(i-\gamma)\partial_t\phi = -\frac{1}{2}\partial^2_x\phi + \varepsilon\phi + g |\phi|^2\phi + \vartheta(x,t),
\label{gp_noise}
\end{equation} 
where noise, coming from a thermal cloud, satisfies 
\begin{equation}
\langle\vartheta(x,t)\vartheta^*(x',t')\rangle= 2\gamma T \delta(x-x')\delta(t-t').
\label{noise_corr}
\end{equation}
Above $\gamma$ represents  
damping 
coming from  thermal cloud -- condensate interactions, $T$ is the thermal cloud temperature,
and $\varepsilon=-\mu$, where $\mu$ is the chemical potential.

Forgetting for a while about damping and noise we see that the system is 
described by the energy functional 
\begin{equation}
{\cal E} = \int dx \ \frac{1}{2}|\partial_x\phi|^2 + V(|\phi|), \ \ \ 
V(|\phi|) = \varepsilon|\phi|^2 + \frac{g}{2}|\phi|^4.
\end{equation}
For $\varepsilon>0$ ($\mu<0$) the minimum of $V(|\phi|)$ corresponds to $\phi=0$: order
parameter vanishes  above condensation temperature, and the system is in the 
symmetric phase. When $\varepsilon<0$ ($\mu>0$) there is a minimum of $V(|\phi|)$ 
at $|\phi|^2 =-\frac{\varepsilon}{g}$, 
and the system is in the broken-symmetry phase where 
$
\phi = \sqrt{-\varepsilon/g}\exp(i\theta)
$: a condensate forms.

We quench the 
system from the symmetric to the broken-symmetry phase. The phase $\theta$ will
not be chosen uniformly in the broken-symmetry phase because the quench 
freezes a characteristic coherence
length, $\hat\xi$, when the system goes out of equilibrium on the symmetric side. This
leads to phase gradients that seed 
soliton-like density notches around which the phase of the 
order parameter changes abruptly \cite{soliton_reference,sengstock}.

For simplicity, we model cooling of a bosonic gas
by assuming that chemical 
potential grows across the critical point, while
damping $\gamma$ and noise correlations remain  
constant. 
This approximation is motivated by the fact that 
defect production takes place  near the critical point, where
chemical potential changes drive the transition 
influencing dynamics most, while the variations in $\gamma$
and noise correlators are of secondary importance.

In the following we  need freeze out values of 
$\hat\varepsilon$ and $\hat\xi$.
As $z=2$ and $\nu=1/2$ for SGPE, Eqs. (\ref{epsilon_hat}) and (\ref{xi_hat}) imply
\begin{equation}
\hat\varepsilon\sim\tau_Q^{-1/2}, \ \hat\xi\sim\tau_Q^{1/4}.
\label{hats}
\end{equation}
We drive the system from the symmetric phase to the broken-symmetry phase
at a constant rate:
\begin{equation}
\varepsilon(t) = -\frac{t}{\tau_Q},
\label{epsilon_t}
\end{equation}
where $t=-\varepsilon_0\tau_Q\to\varepsilon_0\tau_Q$. The evolution 
starts away from the critical point at  $\varepsilon=\varepsilon_0\gg1$.  
The system passes
through the critical point at $\varepsilon,t=0$, and ends its evolution far away from the
critical point at $\varepsilon=-\varepsilon_0$.

\begin{figure}[t]
\includegraphics[width=\columnwidth,clip=true]{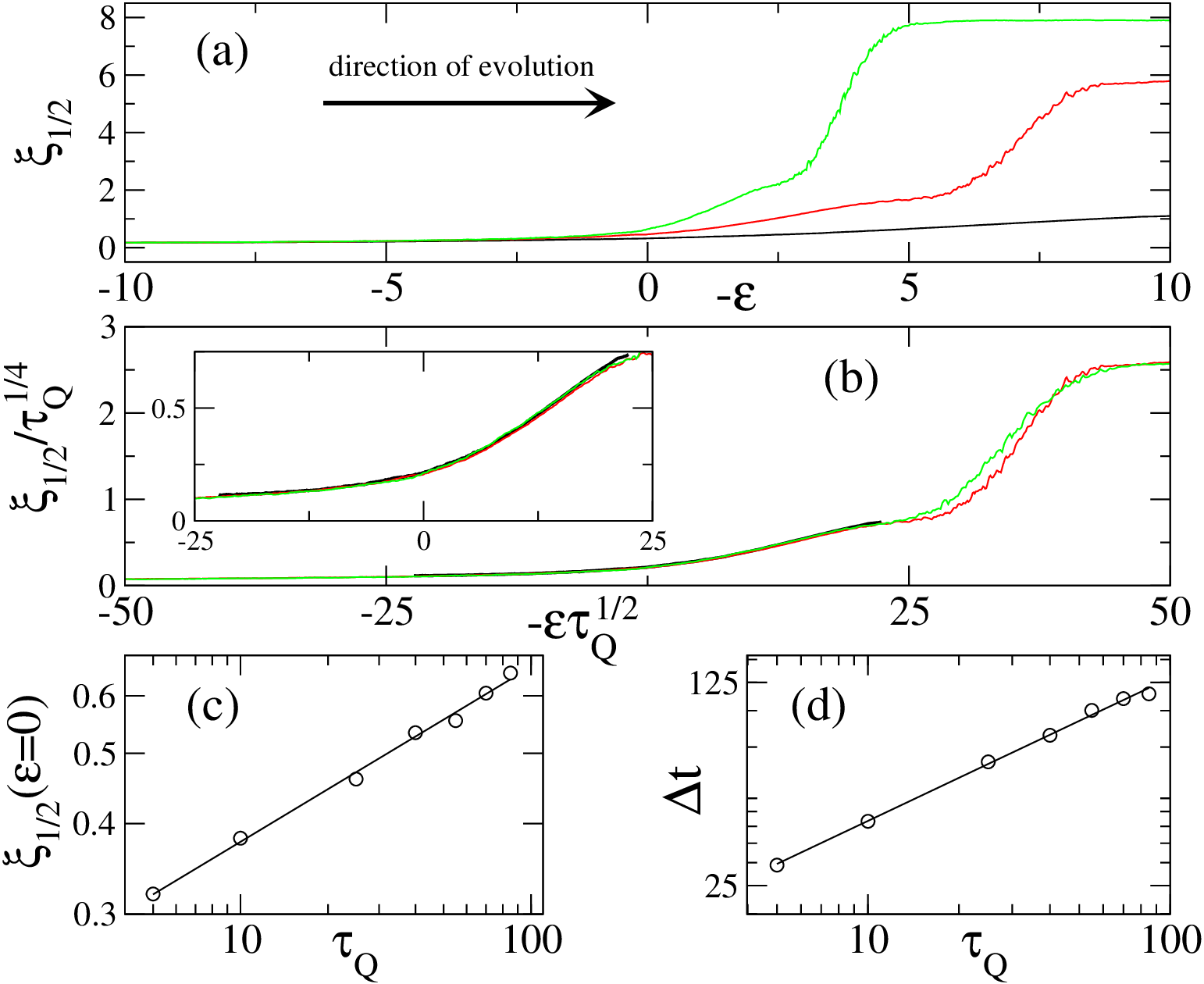}
\caption{Half-width of the averaged two-point correlation 
function (\ref{averaged_C}) during the quench: $\xi_{1/2}$.
{\bf (a) and (b)}: $\xi_{1/2}$ before and after rescalings.
On both plots black, red, and green lines correspond
to $\tau_Q=5, 25$ and $85$, respectively. 
The inset magnifies the region
where the  KZM works best. 
{\bf (c)} Illustration that quench imprints a characteristic 
    length scale $\hat\xi\sim\tau_Q^{\nu/(1+z\nu)}=\tau_Q^{1/4}$ (\ref{hats}).
	We study  $\xi_{1/2}$ at the critical point.
	The scale is logarithmic on both axes: circles show 
	numerics, while the straight line is a fit to 
	numerics: $\ln\xi_{1/2}(\varepsilon=0)=-1.53\pm0.03+(0.24\pm0.01)\ln\tau_Q$.
	The fit  confirms the predicted $1/4$ exponent.
{\bf (d)} Illustration that quench imprints  a characteristic 
          time scale 
		  $\hat t \sim \tau_Q^{z\nu/(1+z\nu)}=\tau_Q^{1/2}$.
		  We study  time $\Delta t$ passing
		  from  entering the broken-symmetry phase
		  to the moment when $\xi_{1/2}/\hat\xi$ reaches a threshold,
		  assumed here to be $0.5$, i.e., to be near 
		  the center of the region where rescalings work on the 
		  broken-symmetry side [see inset of Fig. (b)].
		  The scale is logarithmic on both axes: circles show numerics,
		  while straight line is a fit to numerics: 
		  $\ln\Delta t=2.60\pm0.04+(0.49\pm0.01)\ln\tau_Q$.
		  This is in good agreement with theoretical prediction 
		  of $\Delta t\sim\tau_Q^{1/2}$. 
		  We use for this plot $\hat\xi$ taken from the fit done in 
		  plot (c): $\hat\xi=\tau_Q^{0.24}$.
		  Fits from plots (c) and (d) 
          {\it allow} for  measurement of $z$ and $\nu$ critical 
          exponents. 
          All results
          presented above are  averages  over $100$ runs with different noise.
          See \cite{numerics} for other parameters.
}
\label{width}
\end{figure}

The discussion of KZM  can be analytically illustrated in our model on the 
symmetric side.  
After dropping the nonlinear term we are left with Langevin-type equation:
\begin{equation}
(i-\gamma)\partial_t\phi = -\frac{1}{2}\partial^2_x\phi + \varepsilon\phi + \vartheta(x,t).
\label{nint_noise}
\end{equation}
This equation can be solved exactly providing the following  relaxation time and coherence length 
{\it in equilibrium}:
\begin{equation}
\tau = \frac{1+\gamma^2}{\gamma}\frac{1}{\varepsilon}, \ \xi = \frac{1}{\sqrt{2}}\frac{1}{\sqrt{\varepsilon}},
\label{free}
\end{equation}
respectively. It implies $z=2$ and $\nu=1/2$.
We obtained (\ref{free}) by studying the two point correlation function
\begin{equation}
C(x,t|x',t') = \langle\phi(x,t)\phi^*(x',t')\rangle - \langle\phi(x,t)\rangle \langle\phi^*(x',t')\rangle,
\label{Cxxtt}
\end{equation}
where $\langle\dots\rangle$ denotes averaging over different noise
realizations. We found that  $C(x,t|x',t)$ decays on the length scale 
$\xi$ while $C(x,t|x,t')$ decays on the 
timescale $\tau$.

Solving (\ref{zurek_eq}) with (\ref{epsilon_t}) and (\ref{free})  
we determine that the adiabatic - impulse border is at 
$$
\hat\varepsilon=\sqrt{\frac{1+\gamma^2}{\gamma}}\frac{1}{\sqrt{\tau_Q}},
$$
so the system ``freezes'' when its coherence length equals 
$$
\hat\xi=\xi(\hat\varepsilon)=\frac{1}{\sqrt{2}}\left(\frac{\gamma}{1+\gamma^2}\right)^{1/4}
\tau_Q^{1/4}.
$$ 
Assuming for simplicity that $\varepsilon_0\to\infty$, and solving 
analytically (\ref{nint_noise}) with (\ref{epsilon_t}) we find that far away from 
the critical point the 
system follows adiabatically the instantaneous equilibrium solution 
\begin{equation}
C(x,t|x',t)  = 
\frac{T}{\sqrt{2\varepsilon(t)}}\exp\left(-|x-x'|\sqrt{2\varepsilon(t)}\right).
\label{adiabatic}
\end{equation}

Near the critical point, i.e., when $\varepsilon \lesssim \hat\varepsilon$ we have to
refer to the exact solution that reads
\begin{equation}
C(x,t|x',t) = \langle|\phi(\hat\varepsilon)|^2\rangle_{eq} 
f(|x-x'|/\hat\xi, \varepsilon(t)/\hat\varepsilon),
\label{rescaled_xx}
\end{equation}
where $f(a,b)=\frac{1}{\sqrt{\pi}}\int dk \cos(ka)\exp((b+k^2)^2){\rm
erfc}(b+k^2)$,  $\langle..\rangle_{eq}$ denotes equilibrium averaging, 
and $\langle|\phi(\hat\varepsilon)|^2\rangle_{eq} = T/\sqrt{2\hat\varepsilon}$. Naturally,
 (\ref{rescaled_xx}) reduces
to (\ref{adiabatic}) for $\varepsilon\gg\hat\varepsilon$.

The result (\ref{rescaled_xx}) 
shows that correlations are induced on the length scale $\hat\xi$,
which is given by the correlation length at the border between adiabatic and 
impulse regimes.
Another interesting feature of (\ref{rescaled_xx}) is that it scales as
$\tau_Q^{1/4}$: the equilibrium correlations (\ref{adiabatic}) calculated at 
$\varepsilon(t)=\hat\varepsilon$ scale in the same way. 
Similarly, there is a freeze out time scale $\hat t$ 
imprinted into system's dynamics
through the relation $\varepsilon(t)/\hat\varepsilon = t/\hat t$.
All these results are in perfect agreement with KZM.

On the broken-symmetry side we rely on numerics.
As above, we test our theory by looking at 
the characteristic length and time scales imprinted onto 
the system by the quench. This is done by studying 
half-width of the averaged 
two-point correlation function, i.e.,  
\begin{equation}
\frac{1}{l}\int_0^l dx \left|C(x,t|x+r,t)\right|,
\label{averaged_C}
\end{equation}
where the  averaging over the system size
$l$ is applied. Fig. \ref{width} shows that 
in whole symmetric phase, as well as in the broken-symmetry phase
for $-\varepsilon\tau_Q^{1/2} \lesssim 25$, the following
scaling works very well
\begin{equation}
\xi_{1/2} \sim \hat\xi f(\varepsilon/\hat \varepsilon) 
\sim \tau_Q^{\frac{\nu}{1+z\nu}}f\left(\varepsilon\tau_Q^{\frac{1}{1+z\nu}}\right)=
\tau_Q^{1/4}f\left(\varepsilon\tau_Q^{1/2}\right).
\label{xi_numerics}
\end{equation}

Scaling (\ref{xi_numerics}) is  in perfect agreement with KZM and allows for 
measurement of critical exponents $z$ and $\nu$ (see Fig. \ref{width}). 
We use here SGPE mean-field exponents ($z=2$ and $\nu=1/2$) to compare 
our numerics to theory. 
In an actual experiment, however, other values of critical exponents 
may be relevant (see \cite{esslinger} and references cited therein). 
Indeed, critical exponents 
can be experimentally studied, for example, 
with the help of cavity-assisted cold atom counting recently explored in the 
Esslinger's group at ETH Zurich \cite{esslinger}. There, the system was near equilibrium 
and so the study of two-point correlation functions was aimed at determination
of the critical exponent $\nu$. Nonequilibrium version of this experiment
can provide a new way for measurement of the exponent $\nu$ and can reveal 
the dynamical exponent $z$. All this is possible due to 
presence of the characteristic length scale $\xi_{1/2}$ in the 
non-equilibrium state of the condensate. 
The most striking consequence of existence  of this 
length scale is the creation of solitons. 

\begin{figure}[t]
\includegraphics[width=\columnwidth,clip=true]{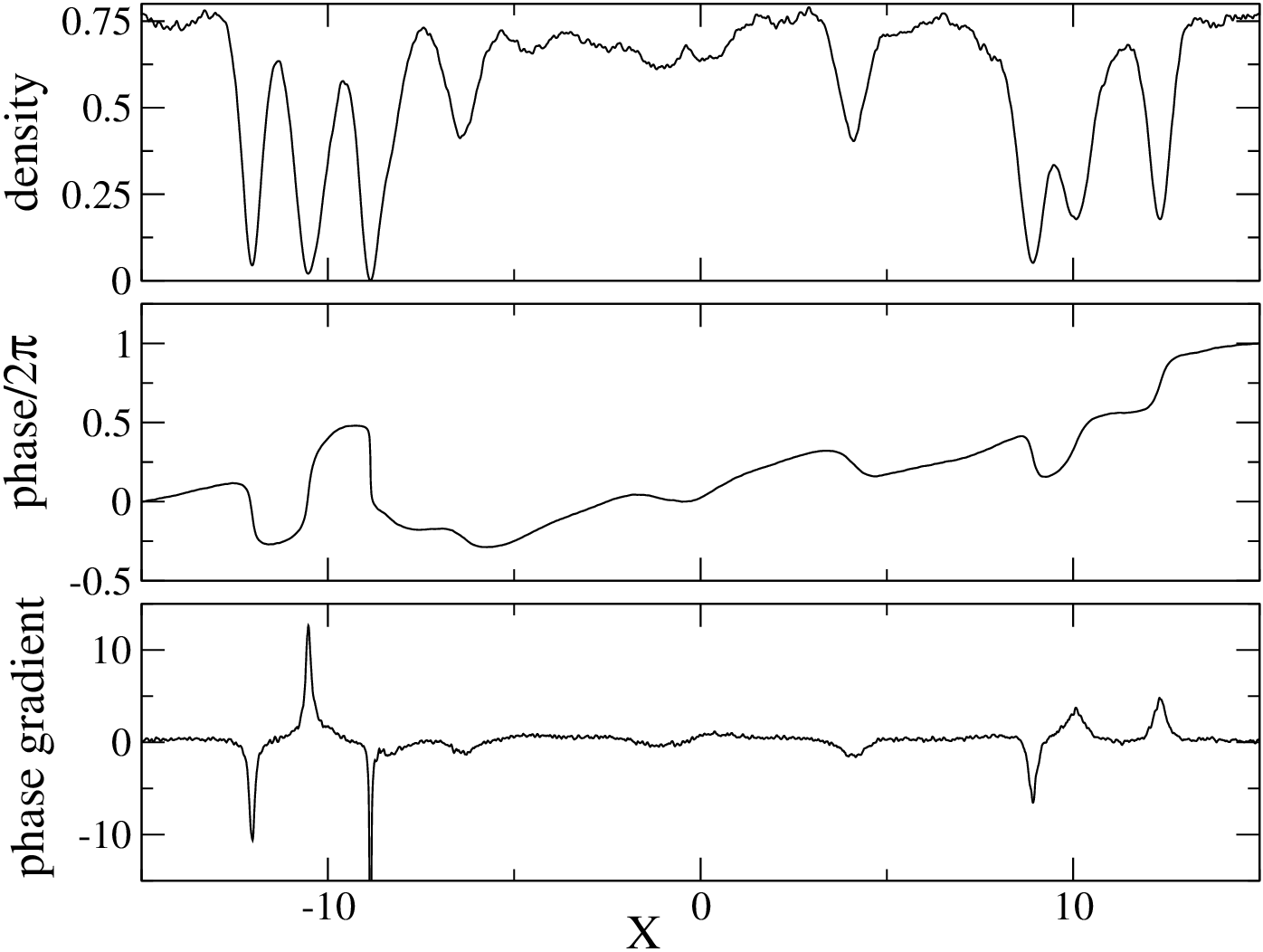}
\caption{Snapshot of density $|\phi(x)|^2$, phase ${\rm arg}(\phi)$, 
and gradient of phase $\frac{d}{dx}{\rm arg}(\phi)$ (the velocity field). 
The snapshot is taken
for one evolution (no averaging) at $\varepsilon(t)=10$ for $\tau_Q=10$.
See \cite{numerics} for other parameters.
}
\label{zdjecie}
\end{figure}

Qualitatively, 
we observe soliton-like solutions in the broken-symmetry phase: see 
Fig. \ref{zdjecie}. There are several deep 
density notches there, and the phase of the order parameter changes steeply 
around them. These are typical signatures of  solitons
\cite{soliton_reference}. They are 
in qualitative agreement with Engels' experiments 
studying density profiles after the condensation process \cite{engels}.

Quantitatively, we would like to find out 
if the critical scalings (\ref{xi_tau_eq}) can be retrieved from 
soliton counting. 
This is of both fundamental and practical interest: 
soliton counting can be significantly easier than measurement of 
the correlation functions. 
The typical number of solitons shall be inversely proportional to the size 
of correlated domains that have chosen to break the symmetry in the 
same way: phase jumps between the domains provide seeds for 
solitons. Thus, we predict  
\begin{equation}
{\rm \# \ of \ solitons} \sim \xi_{1/2}^{-1} 
\sim \tau_Q^{\frac{-\nu}{1+z\nu}}\tilde f\left(\varepsilon\tau_Q^{\frac{1}{1+z\nu}}\right)=
\tau_Q^{-1/4}\tilde f\left(\varepsilon\tau_Q^{1/2}\right).
\label{n_of_solitons}
\end{equation}

\begin{figure}[t]
\includegraphics[width=0.99\columnwidth,clip=true]{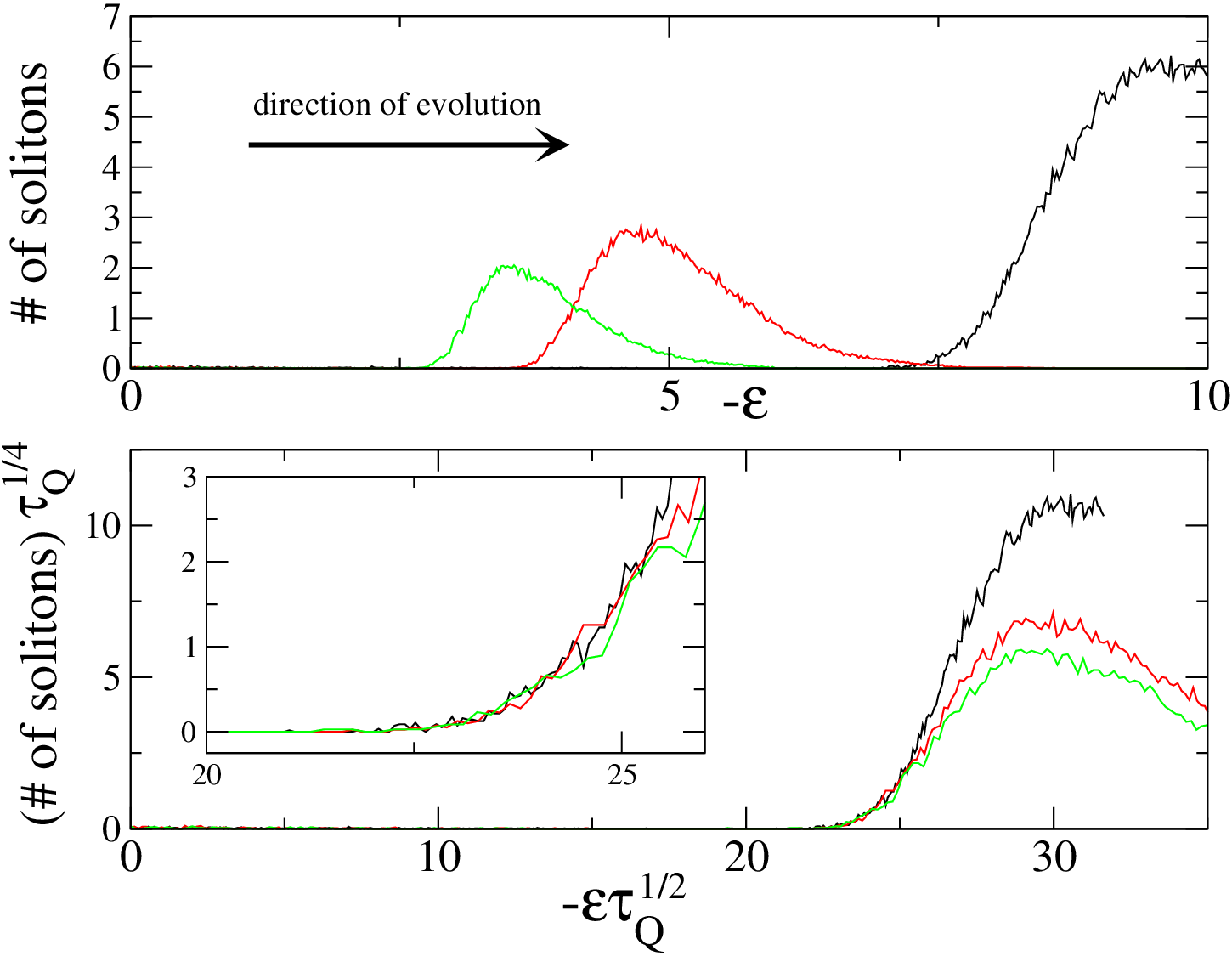}
\caption{Number of solitons in the broken-symmetry phase before and after
rescalings. On both plots black, red and green lines correspond to 
$\tau_Q=10$, $40$ and $70$, respectively. The inset magnifies the region 
where KZM works best.
The results
present average over $100$ runs with different noise.
See \cite{numerics} for other parameters.
}
\label{sol}
\end{figure}

We count solitons by fitting the solitonic solution of
the Gross-Pitaevskii equation (no noise/damping) around every minimum 
of the density $|\phi|^2$:
$$
n_{min} + (n_0-n_{min})\tanh^2[(x-x_0)\sqrt{g(n_0-n_{min})}].
$$
We use only modulus of the order parameter as it is typically the 
only quantity measurable in a standard BEC setup.
The fit gives the three parameters of the soliton
solution: position ($x_0$), minimum density ($n_{min}$), and background density ($n_0$).
The latter two
are related to soliton and sound velocities, respectively. 
We then 
compare the fitted depth of the soliton, $n_0-n_{min}$, to the numerical 
data.
When the two do not differ by more than $50\%$, we count the density 
minimum as a soliton.  We have checked that the same conclusions 
are obtained for other reasonable thresholds between $30\%$ and $60\%$.
The outcome of this procedure is presented in  
Fig. \ref{sol}, where  the number of solitons for 
$-\varepsilon\tau_Q^{1/2} \lesssim 25$ follows closely 
(\ref{n_of_solitons}) 
in agreement with KZM and our studies of  the width of the two-point 
correlation functions. 
The drawback with respect to the latter is that solitons need some 
time to develop before they can be counted, thus their macroscopic 
number can be observed in a narrow window of $20\lesssim-\varepsilon\tau_Q^{1/2}\lesssim25$.
Another complication is that solitons are more vulnerable to damping
than two-point correlation function: 
damping removes solitons from the system by decreasing their depth.
This shall not be a problem with an actual  experiment as there is negligible
damping/noise by the end of condensation \cite{sengstock}.

Our results were obtained for atoms in a
quasi-1D {\it homogeneous} configuration that can be experimentally 
realized \cite{raizen}. 
In harmonic traps with cigar-shaped BEC's homogeneous scalings would be
modified
by causality-related considerations \cite{harm_sol}. 
In either case, the experimental data 
will shed light on the real critical exponents of the cold atom system
and facilitate the first experimental determination of the dynamical, i.e., 
$z$ exponent of the interacting Bose-Einstein condensate. 

Summarizing, we have shown that two-point correlation functions
after non-equilibrium quench 
encode  critical exponents of the system. 
We have also predicted  that 
non-equilibrium condensation results in 
creation of soliton-like excitations whose number depends
on a quench rate via the critical exponents. 
Our results can be ``inverted'' and  used for the  experimental 
determination of the critical exponents for the normal gas -- Bose-Einstein 
condensate phase transition.

We are grateful to Peter Engels for showing us his unpublished experimental data and for 
stimulating discussions. We thank Ashton Bradley for his 
very useful comments.
We acknowledge the support of the U.S. 
Department of Energy through the LANL/LDRD Program.


\begin{thebibliography}{99}
\bibitem{sgpe} The rescaling $t\to t(1+\gamma^2)$ transforms 
(\ref{gp_noise}) to a canonical form of the SGPE: 
$i\partial_t\phi = (1-i\gamma)(-\frac{1}{2}\partial^2_x 
+ g |\phi|^2 -\mu)\phi + \eta$, where
chemical potential $\mu=-\varepsilon$ and noise $\eta$ has the same correlator
as in (\ref{noise_corr}). 
See \cite{ashton}
for  review of SGPE.
\end{thebibliography}
\end{document}